\documentclass{JHEP3} 

\usepackage{epsfig,multicol,bbm,amsmath}

\newcommand{\bn}{\begin{equation}}
\newcommand{\en}{\end{equation}}
\newcommand{\by}{\begin{eqnarray}}
\newcommand{\ey}{\end{eqnarray}}

\newcommand{\de}{\delta}
\newcommand{\ep}{\epsilon}

\newcommand\fverb{\setbox\fverbbox=\hbox\bgroup\verb}
\newcommand\fverbdo{\egroup\medskip\noindent%
            \fbox{\unhbox\fverbbox}\ }
\newcommand\fverbit{\egroup\item[\fbox{\unhbox\fverbbox}]}
\newbox\fverbbox

\newcommand{\nablaslash}{\not{\hbox{\kern-3pt $\nabla$}}}

\title{Three-dimensional $\mathcal N=8$ conformal supergravity and its coupling to BLG M2-branes}

\author{Ulf Gran and Bengt E.W.~Nilsson
\\\\
Fundamental Physics\\
Chalmers University of Technology\\
SE-412 96 G\"oteborg, Sweden\\

{\tt {\footnotesize ulf.gran@chalmers.se, tfebn@chalmers.se}}}

\abstract{This paper is concerned with the problem of coupling the
$\mathcal N=8$ superconformal Bagger-Lambert-Gustavsson (BLG) theory
to $\mathcal N=8$ conformal  supergravity in three dimensions. We
start by constructing the on-shell $\mathcal N=8$ conformal
supergravity in three dimensions consisting of a Chern-Simons type
term for each of the gauge fields: the spin connection, the $SO(8)$
R-symmetry gauge field and the spin 3/2 Rarita-Schwinger (gravitino)
field. We then proceed to couple this theory to the BLG theory. The
final theory should have the same physical content, i.e., degrees of
freedom, as the ordinary BLG theory. We discuss briefly the
properties of this "topologically gauged" BLG theory and why this
theory may be useful.}


\begin{document}



\section{Introduction}

Recently a basically unique three-dimensional maximally ($\mathcal
N=8$) superconformal theory was constructed by Bagger and Lambert,
and by Gustavsson (BLG)
\cite{Bagger:2006sk,Gustavsson:2007vu,Bagger:2007jr,Bagger:2007vi}.
It is the purpose of this paper to develop the corresponding
($\mathcal N=8$) conformal supergravity theory and couple it to the
BLG theory.

 The BLG theory, containing a Chern-Simons  gauge field
  coupled to matter fields,
 was originally proposed to describe multiple M2-branes.
An interesting aspect of the fact that the BLG theory is a
Chern-Simons theory \cite{Schwarz:2004yj} is its potential
importance also in the context of condensed matter applications. The
multiple M2-brane interpretation has, however, met with a number of
problems related to the algebraic structure of the theory. The BLG
construction is based on a four-index structure constant for a
three-algebra with a Euclidean metric. This three-algebra is known
\cite{Papadopoulos:2008sk,Gauntlett:2008uf} to have basically only
one realization, $\mathcal A_4$, related to the ordinary Lie algebra
$so(4)$. This seems to be limiting the role of the BLG theory to
stacks of two M2-branes
\cite{Lambert:2008et,Distler:2008mk,Aharony:2008ug}.

It may be of some interest to couple the BLG theory to supergravity.
In fact, in the context of $AdS_5/CFT_4$, similar couplings of a
superconformal field theory to its supergravity counterpart have
been considered in the past, see, e.g.,
\cite{Liu:1998bu}\footnote{We are grateful to Arkady Tseytlin for
discussions on this point.} and references therein. A coupling to
supergravity also provides a framework for curved M2 branes and may
perhaps be used in a way similar to how quantum properties of the
string are usually defined. (This may be more natural in the context
of the ABJM $\mathcal N=6$ theory \cite{Aharony:2008ug} which can
describe one as well as many M2 branes\footnote{For several reasons
one may, in fact, suspect that globally  there is no distinction
between one and several M2 branes.}. This theory can most likely be
coupled to conformal supergravity in the same way as done here for
the BLG theory.) The geometric description of the superstring
coupled to supergravity, generally referred to as the Polyakov
string, was first given in \cite{Brink:1976sc} and later used by
Polyakov \cite{Polyakov:1981re} to define the string at the quantum
level. There is, of course, an interesting issue at this stage for
such an interpretation to be viable in the M2 case, related to the
fact that BLG/ABJM type theories appear to be in a static gauge of a
would-be covariant theory. We have no comments about this at the
moment and regard this work only as a possible step in this
direction.

However, for the coupling to gravity to make sense in this latter
context, the coupled three-dimensional BLG (or ABJM) theory should
not pick up any new propagating degrees of freedom. Thus the
supergravity theory needs to be special, in some sense topological
before being coupled to matter. For $\mathcal N=1$ there is such a
theory in three dimensions as shown by Deser and Kay in
\cite{Deser:1982sw}. It consists of two Chern-Simons type terms, one
for the spin connection and one for its superpartner the
Rarita-Schwinger field. Although none of them are conventional
Chern-Simons terms (e.g., the spin connection is constructed from
the dreibein), we will refer to both as Chern-Simons terms. Some
issues related to the physical content of theories of this kind have
been addressed in \cite{Horne:1988jf,Guralnik:2003we}. For instance,
the equation of motion for the dreibein in the pure gravity case
restricts the geometry to be conformally flat \cite{Horne:1988jf}.

In this paper we construct the $\mathcal N=8$ version of this
supergravity theory which interestingly enough turns out to be
rather simple; starting from the Deser-Kay $\mathcal N=1$ theory
\cite{Deser:1982sw} one just gives the spinors an extra $SO(8)$
spinor index and adds a Chern-Simons term for the corresponding
R-symmetry gauge field. It is then rather straightforward to show
that this theory is invariant under the local symmetries,
supersymmetry, special superconformal, and dilatations. (The
Lagrangian of this theory can also be obtained by starting from the
gauged superconformal algebra and subject it to curvature
constraints as shown in
\cite{VanNieuwenhuizen:1985ff,Lindstrom:1989eg}.) It is then
possible to couple this conformal supergravity theory to the BLG
theory using familiar methods. In this paper we will perform this
coupling up to some higher order interaction terms between the two sectors. The resulting theory will
here sometimes be referred to as the topologically gauged BLG theory
since the global symmetries of the BLG theory, namely Poincar\'e,
$\mathcal N=8$ supersymmetry and $SO(8)$ R-symmetry, are here all
being gauged by the introduction of gauge fields and the
corresponding Chern-Simons terms. The introduction of levels $k$ can
be done separately in the BLG sector
\cite{Lambert:2008et,Distler:2008mk,Aharony:2008ug} and in the
supergravity sector \cite{Horne:1988jf} raising some interesting
questions. This is discussed further in the last section.

 The paper is organized as follows. In section two we construct the
$\mathcal N=8$ conformal (or topological) supergravity by writing
down an on-shell Lagrangian containing only three types of
Chern-Simons terms, one for each gauge symmetry. We then explicitly
demonstrate that this supergravity theory has the required $\mathcal
N=8$ local symmetries. In section three we review the BLG theory and
present in detail the coupling of it to the $\mathcal N=8$ conformal
supergravity given in section two. The last section contains
conclusions and some further comments.

\section{Pure topological ${\mathcal N=8}$ supergravity in three dimensions}
The off-shell field content of three-dimensional ${\mathcal N=8}$
conformal supergravity is
\begin{equation}
e_{\mu}{}^{\alpha}\,\,[0],\,\,\chi_{\mu}^{i}\,\,[-\tfrac{1}{2}],\,\,B_{\mu}^{ij}\,\,[-1],
\,\,b_{ijkl}\,\,[-1],\,\,\rho_{ijk}\,\,[-\tfrac{3}{2}],\,\,c_{ijkl}\,\,[-2],
\end{equation}
where we have given the conformal dimension in brackets after each
field. This set of fields constitute an off-shell multiplet of
$\mathcal N=8$ three-dimensional conformal supergravity
\cite{Howe:1995zm} as indicated by the degree of freedom count
(which is just as in four dimensions but then on-shell). The task
now is to construct a topological Lagrangian from a set of
Chern-Simons terms. In fact, by checking which scale invariant terms
can be constructed from the above set of fields one concludes that
the last three fields will satisfy algebraic field equations. This
means that we can construct the on-shell Lagrangian using only the
three gauge fields of 'spin' 2, 3/2 and 1, i.e.
$e_{\mu}{}^{\alpha}[0],\,\,\chi_{\mu}^{i}[-\tfrac{1}{2}],\,\,B_{\mu}^{ij}[-1]$.
(Note that the $i$ index used here corresponds in the following to
the $SO(8)$ spinor index that is not explicitly written out for the
supersymmetry parameter. The R-symmetry gauge field in the adjoint
of $SO(8)$ may, due to triality, be given a pair of antisymmetric
indices in any of the three eight-dimensional representations.)

Inspired by the work of Deser and Kay \cite{Deser:1982sw}, van
Nieuwenhuizen \cite{VanNieuwenhuizen:1985ff}, and Lindstr\"om and
Ro\v cek \cite{Lindstrom:1989eg}, we start from a Lagrangian of the
form\footnote{The Lagrangian used here was in fact given in
\cite{Lindstrom:1989eg} based on a generalization of the
superconformal algebra method of \cite{VanNieuwenhuizen:1985ff}. We
will, however, base our discussion entirely on methods related to
those of Deser and Kay in \cite{Deser:1982sw}. }
\begin{eqnarray}
L&=&\frac{1}{2}\epsilon^{\mu\nu\rho}
Tr_{\alpha}(\tilde\omega_{\mu}\partial_{\nu}\tilde\omega_{\rho}+
\frac{2}{3}\tilde\omega_{\mu}\tilde\omega_{\nu}\tilde\omega_{\rho})
-\epsilon^{\mu\nu\rho}Tr_i
(B_{\mu}\partial_{\nu}B_{\rho}+\frac{2}{3}B_{\mu}B_{\nu}B_{\rho})\notag\\[1mm]
&&-i e^{-1} \epsilon^{\alpha\mu\nu}\epsilon^{\beta\rho\sigma}(\tilde
D_{\mu}\bar{\chi}_{\nu}\gamma_{\beta}\gamma_{\alpha}\tilde
D_{\rho}\chi_{\sigma}),
\end{eqnarray}
where $\tilde\omega$ is the spin connection and the traces in the
first and second terms are over the vector representation of the
Lorentz group $SO(1,2)$ and the R-symmetry group $SO(8)$,
represented by indices $\alpha$ and $i$, respectively. Note that the
coefficient in front of the R-symmetry Chern-Simons term may seem
non-standard but as we will see below the $\mathcal N=8$
supersymmetry properties depends crucially on the value of this
coefficient.

We will frequently use the standard notation \cite{Deser:1982sw}
\begin{equation}
f^{\mu}=\frac{1}{2}\epsilon^{\mu\nu\rho}\tilde D_{\nu}{\chi}_{\rho},
\end{equation}
which makes the Rarita-Schwinger term read
\begin{equation}
-4i\bar f^{\mu} \gamma_{\beta}\gamma_{\alpha}
f^{\nu}(e_{\mu}{}^{\alpha}e_{\nu}{}^{\beta}e^{-1}),
\end{equation}
where we have spelt out explicitly all dependence of the dreibein
that needs to be varied when checking supersymmetry.

The standard procedure to obtain local supersymmetry is to start by
adding Rarita-Schwinger terms to the dreibein-compatible $\omega$ in
order  to obtain a supercovariant version of it. That is
\begin{equation}
\tilde\omega_{\mu \alpha\beta}=\omega_{\mu \alpha\beta}+K_{\mu
\alpha \beta},
\end{equation}
where
\begin{equation}
\omega_{\mu\alpha\beta}=\frac{1}{2}(\Omega_{\mu\alpha\beta}-\Omega_{\alpha\beta\mu}+\Omega_{\beta\mu\alpha}),
\end{equation}
with
\begin{equation}
\Omega_{\mu\nu}{}^{\alpha}=\partial_{\mu}e_{\nu}{}^{\alpha}-\partial_{\nu}e_{\mu}{}^{\alpha},
\end{equation}
and
\begin{equation}
K_{\mu\alpha\beta}=-\frac{i}{2}(\bar\chi_{\mu}\gamma_{\beta}\chi_{\alpha}-
\bar\chi_{\mu}\gamma_{\alpha}\chi_{\beta}-\bar\chi_{\alpha}\gamma_{\mu}\chi_{\beta}).
\end{equation}
This combination of spin connection and contorsion is
supercovariant, i.e., derivatives on the supersymmetry parameter
cancel out if $\tilde\omega_{\mu \alpha\beta}$ is varied under the
ordinary transformations of the dreibein and Rarita-Schwinger field:
\begin{equation}
\delta e_{\mu}{}^{\alpha}=i\bar\epsilon \gamma^{\alpha}\chi_{\mu},
\,\,\, \delta\chi_{\mu}= \tilde D_{\mu}\epsilon.
\end{equation}

The covariant derivative appearing in the Lagrangian and in the
variation of the Rarita-Schwinger field takes the following form
acting on a spinor
\begin{equation}
\tilde
D_{\mu}\epsilon=\partial_{\mu}\epsilon+\frac{1}{4}\tilde\omega_{\mu\alpha
\beta}\gamma^{\alpha \beta}\epsilon+ \frac{1}{4}B_{\mu
ij}\Gamma^{ij}\epsilon,
\end{equation}
that is, both the Lorentz $SO(1,2)$ and the R-symmetry $SO(8)$
groups are gauged. Note that the spinors in the gravity sector,
i.e., the susy parameter and the Rarita-Schwinger field, are of the
same $SO(8)$ chirality while the spinor in the BLG theory is of
opposite chirality.

Our goal now is to show that the above Lagrangian is $\mathcal N=8$
supersymmetric (up to a total divergence) under the above
transformations of the dreibein and the Rarita-Schwinger field
together with a transformation of the $SO(8)$ R-symmetry gauge field
$B_{\mu ij}$ that will be determined in the course of the
calculation. This superconformal  $\mathcal N=8$ supergravity theory
will then be coupled to the BLG theory in the next section.

We will derive the variation of the Lagrangian following closely the
steps in the $\mathcal N=1$ case given by Deser and Kay in
\cite{Deser:1982sw}. There is, however, good reason to be somewhat
more explicit than in that paper since we do it for $\mathcal N=8$
and will need to spell out in detail where the two
 calculations differ. Our derivation will make use of a Fierz basis (see the appendix)
 which will turn out to
 simplify the calculations quite a bit.

 Introducing the dual $SO(8)$ R-symmetry and curvature fields (see \cite{Deser:1982sw})
 \begin{equation}
G^{*\mu}_{ij}=\frac{1}{2}\epsilon^{\mu\nu\rho}G_{\nu \rho ij},\,\,\,
\tilde
R^{*\mu}{}_{\alpha\beta}=\frac{1}{2}\epsilon^{\mu\nu\rho}\tilde
R_{\nu\rho\alpha\beta}
\end{equation}
and similarly for $\tilde\omega$, as well as the double and triple
duals
\begin{equation}
\tilde
R^{**\mu,\alpha}=\frac{1}{2}\epsilon^{\alpha\beta\gamma}\tilde
R^{*\mu}{}_{\beta\gamma}, \,\,\, \tilde
R^{***}_{\mu}=\frac{1}{2}\epsilon_{\mu\nu\alpha}\tilde
R^{**\nu,\alpha},
\end{equation}
where in the last expression only the contorsion part of the Riemann tensor contributes. In fact, one can show that
\begin{equation}
\tilde{R}^{***}_{\mu}=i e^2 \bar\chi_\nu \gamma_{\mu}f^{\nu}.
\end{equation}

From the fact that the affine connection and spin connection are
related by
\begin{equation}
\Gamma_{\mu\nu}^{\rho}=\omega_{\mu}{}^{\alpha}{}_{\beta}e_{\nu}{}^{\beta}e_{\alpha}{}^{\rho}+
e_{\alpha}{}^{\rho}\partial_{\mu} e_{\nu}{}^{\alpha},
\end{equation}
and that the variation of the affine connection is
\begin{equation}
\delta\Gamma_{\mu\nu}^{\rho}=\frac{1}{2}g^{\rho\sigma}(D_{\mu}\delta
g_{\nu\sigma}+D_{\nu}\delta g_{\mu\sigma}-D_{\sigma}\delta
g_{\mu\nu}),
\end{equation}
we find directly that
\begin{equation}
\delta\tilde\omega^{*\alpha}_{\mu}=-2i(\bar\epsilon\gamma_{\mu}f^{\alpha}-
\frac{1}{2}e_{\mu}{}^{\alpha}\bar\epsilon\gamma_{\nu}f^{\nu}).
\end{equation}

Combining this result with the fact that the commutator of two
supercovariant derivatives, acting on a spinor, is
\begin{equation}
[\tilde D_{\mu},\tilde D_{\nu}]=\frac{1}{4}\tilde
R_{\mu\nu\alpha\beta}\gamma^{\alpha\beta}+
 \frac{1}{4}G_{\mu\nu ij}\Gamma^{ij},
\end{equation}
we find that  the symmetric part of $R^{**\mu,\alpha}$ cancels in
the supersymmetry variation of the dreibein and gravitino
Chern-Simons terms. Performing also the variation of the
Chern-Simons term for the $SO(8)$ gauge field we find that also
$G^{*\mu}_{ij}$ cancels provided we choose the variation of $B_{\mu
ij}$ to be
\begin{equation}
\delta B_{\mu}^{ij}=-\frac{i}{2}e^{-1}\bar\epsilon\Gamma^{ij}\gamma_{\nu}\gamma_{\mu}f^{\nu}.
\end{equation}

Inserting these variations into $\delta L$ gives
\begin{eqnarray}
\delta L&=&\delta L_1 + \delta L_2 + \delta L_3 + \delta L_4,\notag\\[1mm]
\delta L_1&=&
4\bar\epsilon(\gamma_{\alpha}\gamma_{\beta}f^{\alpha})\bar
f^{\mu}\gamma^{\beta}\chi_{\mu},\cr \delta L_2&=&8\bar
f^{\mu}(\gamma_{\alpha}\gamma_{\beta}f^{\alpha})(\bar\epsilon\gamma^{\beta}\chi_{\mu}
-\frac{1}{2}e_{\mu}{}^{\beta}\bar\epsilon\gamma^{\nu}\chi_{\nu}),
\cr \delta L_3&=&4(\bar
f^{\alpha}\gamma_{\beta}\gamma_{\alpha})\gamma_{\gamma}\chi_{\mu}\epsilon^{\beta\mu\nu}
(\bar\epsilon\gamma_{\nu}f^{\gamma}-\frac{1}{2}e_{\nu}{}^{\gamma}\bar\epsilon\gamma^{\rho}f_{\rho}),\cr
\delta L_4&=&-\frac{1}{2}(\bar
f^{\alpha}\gamma_{\beta}\gamma_{\alpha})\Gamma^{ij}\chi_{\mu}\epsilon^{\mu\beta\gamma}\bar\epsilon
\Gamma_{ij}(\gamma_{\delta}\gamma_{\gamma}f^{\delta}).
\end{eqnarray}

In order to show that the variation of the Lagrangian vanishes some
of the terms in the above expression must be rearranged by Fierz
transformations. As we will see later it will turn out to be
convenient to review the $\mathcal N=1$ case before turning to the
more complicated case of $\mathcal N=8$. To proceed in a systematic
manner we have chosen to pick a basis of $\mathcal N=1$ expressions
consisting of $(\bar f...f)(\bar\epsilon....\chi)$ where the dots
correspond to either a charge conjugation matrix or such a matrix
times a three dimensional gamma matrix (recall that all the spinors
are Majorana). An independent set of such expressions is defined in
the Appendix.

By applying the Fierz transformations to $\delta L_1$ and $\delta
L_3$ above and expressing all terms so obtained in the Fierz basis
one can show, after some ${\mathcal N=1}$ Fierz calculations, that
they exactly cancel $\delta L_2$. This is the result of Deser and
Kay \cite{Deser:1982sw}.

It now becomes rather easy to establish that also for $\mathcal N=8$
the variation will vanish when $\delta L_4$ is included and use is
made of the full $\mathcal N=8$ Fierz identity for $SO(8)$ spinors
of the same chirality, i.e.,
\begin{eqnarray}
\bar A B \bar C D& =&-\frac{1}{16}(\bar A D \bar C B + \bar A
\gamma_{\alpha}D \bar C \gamma_{\alpha}B \cr &&-\frac{1}{2}\bar A
\Gamma^{ij}D \bar C \Gamma^{ij} B- \frac{1}{2}\bar A\gamma_{\alpha}
\Gamma^{ij} D \bar C\gamma_{\alpha} \Gamma^{ij} B\cr
&&+\frac{1}{48}\bar A  \Gamma^{ijkl}D \bar C  \Gamma^{ijkl}B+
\frac{1}{48}\bar A \gamma_{\alpha} \Gamma^{ijkl}D \bar
C\gamma_{\alpha}  \Gamma^{ijkl}B).
\end{eqnarray}

The argument is as follows. From the $\mathcal N=1$ case, for
instance by using the basis given in the Appendix and the $\mathcal
N=1$ Fierz identity
\begin{eqnarray}
\bar A B \bar C D =-\frac{1}{2}(\bar A D \bar C B + \bar A
\gamma_{\alpha}D \bar C \gamma_{\alpha}B),
\end{eqnarray}
we conclude that after Fierzing $\delta L_1+ \delta L_3=-\delta
L_2$. This means that in the $\mathcal N=8$ case we have instead
that $\delta L_1+ \delta L_3=-\frac{1}{8}\delta L_2$ and we are
missing $\frac{7}{8}\delta L_2$ which must come from Fierzing
$\delta L_4$.

That this in fact is exactly what happens is most easily seen by
Fierzing $\delta L_4$ keeping the factors
$\gamma_{\alpha}\gamma_{\beta}f^{\alpha}$ intact and collecting the
$\Gamma^{ij}$ in the same factor. The result of the Fierzing is
\begin{eqnarray}
(\bar f^{\alpha}\gamma_{\beta}\gamma_{\alpha})
\gamma^{\mu}\epsilon^{\nu \beta \delta}
(\gamma_{\gamma}\gamma_{\delta}f^{\gamma})
\bar\epsilon\gamma_{\mu}\chi_{\nu}=4\bar
f^{\mu}(\gamma_{\alpha}\gamma_{\beta}f^{\alpha})(\bar\epsilon\gamma^{\beta}\chi_{\mu}
-\frac{1}{2}e_{\mu}{}^{\beta}\bar\epsilon\gamma^{\sigma}\chi_{\sigma}),
\end{eqnarray}
where the right hand side has been derived by writing $\gamma^{\nu
\beta \delta}$ instead of $\epsilon^{\nu \beta \delta}$ and then
multiplying in the explicit $\gamma^{\mu}$ into it.

Turning finally to the Fierz terms containing $\Gamma_{ij}$ and
$\Gamma_{ijkl}$, the latter terms  can be seen to cancel directly
using the same Fierz relations as for the terms without any
$\Gamma$'s. The cancelation of the $\Gamma_{ij}$ does however
require a separate calculation using the second basis set in the
Appendix. This cancelation has also been verified proving that the
theory has $\mathcal N=8$ local supersymmetry.

We have also explicitly verified that the theory constructed here is
locally scale invariant (denoted by an index $\Delta$) and possesses
$\mathcal N=8$ superconformal (shift) symmetry (denoted by $S$) with
the following transformation rules (where $\phi$ is the local scale
parameter and $\eta$ the local shift parameter)
\begin{eqnarray}
\delta_{\Delta} e_{\mu}{}^{\alpha}&=&-\phi(x)e_{\mu}{}^{\alpha},\notag\\[1mm]
\delta_{\Delta} \chi_{\mu}&=&-\tfrac{1}{2}\phi(x)\chi_{\mu},\notag\\[1mm]
\delta_{\Delta} B_{\mu}^{ij}&=&0,
\end{eqnarray}
and
\begin{eqnarray}
\delta_S e_{\mu}{}^{\alpha}&=& 0,\notag\\[1mm]
\delta_S \chi_{\mu}&=&
\gamma_{\mu} \eta, \notag\\[1mm]
 \delta_S B_{\mu}^{ij}&=& \tfrac{i}{2}\bar \eta \Gamma^{ij}
 \chi_{\mu}.
\end{eqnarray}
Verifying invariance under the latter transformations requires Fierz
transformations similar to those used above to demonstrate $\mathcal
N=8$ supersymmetry. The calculations performed here may be
facilitated by the use of the Mathematica package $GAMMA$
\cite{Gran:2001yh}.

\section{The $\mathcal N=8$ gauged BLG theory}

In this section we first review the (ungauged) superconformal matter
sector, i.e., the ordinary BLG theory, to which we then would like
to couple the superconformal gravity derived in the previous
section. The resulting "gauged" BLG theory is derived in the second
subsection up to some higher order interaction terms between the two
sectors.

\subsection{Review of the ungauged $\mathcal N=8$ superconformal BLG}

The BLG theory contains three different fields; the two propagating
ones $X^{i}{}_{a}$ and $\Psi_{a}$, which are three-dimensional
scalars and spinors, respectively, and the auxiliary gauge field
$\tilde{A}_{\mu}{}^{a}{}_{b}$. Here the indices $a,\,b,\,\ldots$ are
connected to the three-algebra and some $n$-dimensional basis
$T^{a}$, while the $i,\,j,\,k,\,\ldots$ indices are $SO(8)$ vector
indices. The spinors transform under a spinor representation of
$SO(8)$ but the corresponding index is not written out explicitly.
Indices $\mu,\,\nu,\, \ldots$ are vector indices on the flat
M2-brane world volume.

Using these fields one can write down $\mathcal N=8$ supersymmetry
transformation rules and  covariant field equations.  This is
possible without introducing a metric on the three-algebra. In such
a situation the position of the indices on the structure constants
is fixed as $f^{abc}{}_d$. The corresponding fundamental identity
needed for supersymmetry and gauge invariance then reads
\cite{Bagger:2006sk,Gustavsson:2007vu,Bagger:2007jr,Bagger:2007vi},
\begin{equation}
f^{abc}{}_g f^{efg}{}_d = 3f^{ef[a}{}_g f^{bc]g}{}_d  \,, \label{FI}
\end{equation}
which can be written in the following alternative but equivalent
form \cite{Gran:2008vi},
\begin{equation}
f^{[abc}{}_g f^{e]fg}{}_d = 0 \,. \label{WFI}
\end{equation}

The construction of a Lagrangian requires the introduction of a
metric on the three-algebra. As discussed above,  if one wants to
describe more general Lie algebras than $so(4)$,  this metric must
be degenerate \cite{Gran:2008vi} or non-degenerate but indefinite
\cite{Gomis:2008uv,Benvenuti:2008bt,Ho:2008ei}. Finally, to
construct an action one also needs to introduce the basic gauge
field $A_{\mu ab}$ \footnote{However, already gauge invariance of
the field equations requires the introduction of this gauge field
\cite{Gran:2008vi}.} which is related to the previously defined
gauge field and structure constants as follows:
\begin{equation}
\tilde{A}_{\mu}{}^{a}{}_b=A_{\mu cd}f^{cda}{}_{b}\,.
\end{equation}

The BLG Lagrangian is
 \cite{Bagger:2007jr}
\begin{eqnarray}
{\cal L} &=& -\tfrac{1}{2}(D_\mu X^{ia})(D^\mu X^i{}_a) +
\tfrac{i}{2} \bar \Psi^a \gamma^\mu D_\mu \Psi_a -
 \tfrac{i}{4} \bar\Psi_b \Gamma_{ij} X^i{}_c X^j{}_d \Psi_a f^{abcd}
\nonumber\\[1mm] &&-V
+\tfrac{1}{2}\varepsilon^{\mu\nu\lambda}\left( f^{abcd}A_{\mu
ab}\partial_\nu A_{\lambda cd} +
 \tfrac{2}{3} f^{cda}{}_g f^{efgb} A_{\mu ab}  A_{\nu cd} A_{\lambda ef}  \right) \,,
\end{eqnarray}
where the potential is given by
\begin{equation}
V = \tfrac{1}{12} f^{abcd}f^{efg}{}_d X^i{}_a X^j{}_b X^k{}_c
X^i{}_e X^j{}_f X^k{}_g\,.
\end{equation}
Note that in terms of $\tilde A$ the Chern-Simons term becomes
\begin{equation}
{\cal L}_{CS}=\tfrac{1}{2}\varepsilon^{\mu\nu\lambda}\left( A_{\mu
ab}\partial_\nu \tilde A_{\lambda}{}^{ab} + \tfrac{2}{3}
A_{\mu}{}^a{}_b \tilde A_{\nu}{}^b{}_c \tilde A_{\lambda}{}^c{}_{a}
\right)
\end{equation}

The BLG transformation rules for (global) $\mathcal N=8$
supersymmetry are
\begin{eqnarray}
\delta X_{i}^a&=& i\epsilon \Gamma_i \Psi^a,\notag\\[1mm]
\delta\Psi_a&=&\bar D_{\mu}X^i_a \gamma^{\mu}\Gamma^i \epsilon
+\tfrac{1}{6}X^i_b X^j_c X^k_d \Gamma^{ijk}\epsilon
f^{bcd}{}_a,\notag\\[1mm] \delta \tilde A_{\mu}{}^a{}_b&=&
i\bar\epsilon \gamma_{\mu} \Gamma^i X_c^i \Psi_d f^{cda}{}_b.
\end{eqnarray}

\subsection{Coupling $\mathcal N=8$ conformal supergravity to BLG matter}

We now turn to the construction of the gauged BLG Lagrangian. The
coupling of the BLG theory to the $\mathcal N=8$ conformal
supergravity theory discussed in the previous section follows from
standard techniques. As a first step in its derivation we restrict
ourselves to terms in the Lagrangian that give rise to
$(cov.der.)^2$ or $(cov.der.)^3$ terms when varied under
supersymmetry and show  that all such terms cancel in $\delta L$.
Including also some other terms, like those that complete the
supercurrent, we will use the following Lagrangian as our starting
point
\begin{eqnarray}
L&=&L_{conf.sugra}+L_{BLG}^{cov}+L_{supercurrent},
\end{eqnarray}
where $L_{conf.sugra}$ was given in section two,
\begin{eqnarray}
L_{BLG}^{cov}=e(-\tfrac{1}{2}g^{\mu\nu}\tilde D_{\mu}X^{ia} \tilde
D_{\nu}X_{ia} +\tfrac{i}{2}\bar\Psi^a
\gamma^{\alpha}e_{\alpha}{}^{\mu}\tilde
D_{\mu}\Psi_a+L_{Yukawa}-V)+L_{CS(A)},
\end{eqnarray}
and
\begin{eqnarray}
 L_{supercurrent}=
Aie \bar\chi_{\mu}\Gamma^i \gamma^{\nu}\gamma^{\mu}\Psi^a (\tilde
 D_{\nu}X^{ia}-\hat A\tfrac{i}{2}\bar\chi_{\nu}\Gamma^i
 \Psi^a)\nonumber
 \end{eqnarray}
\begin{eqnarray}
 +\hat Bie\bar\chi_{\mu}\gamma^{\mu}\Gamma^{ijk}\Psi_a(X^i_bX^j_cX^k_d)f^{abcd}+
 \hat Ce\bar\chi_{\mu}\Gamma^{ijkl}\chi^{\mu}(X^i_aX^j_bX^k_cX_d^l)f^{abcd},
\end{eqnarray}
and then add terms as they become necessary for proving
supersymmetry to the order in covariant derivatives at which we are
working. Here the derivatives are covariant under all local
symmetries of the theory. Note that the hatted parameters $\hat A$,
$\hat B$ and $\hat C$ in the supercurrent are not determined by the
$(\tilde D_{\mu})^2$ calculation. In fact, at the end of this
subsection we will determine also these coefficients by demanding
cancelation of terms that contain fewer derivatives but are of power
four or higher in $X$. The whole Lagrangian is then known up to some
fermion terms without derivatives that might be
needed in addition to the ones already present in the covariant
derivatives. The final step of proving cancelation also of the one-
and non-derivative terms in $\delta L$ is fairly elaborate and will
be presented elsewhere.

The new terms that arise in the computation are the following
\begin{eqnarray}
A'i\epsilon^{\mu\nu\rho}\bar\chi_{\mu}\Gamma^{ij}\chi_{\nu}(X_a^i\tilde
D_{\rho}X_a^j)+ A''i\bar f^{\mu}\Gamma^i\gamma_{\mu}\Psi_a X_a^i,
\end{eqnarray}
together with
\begin{eqnarray}
-\tfrac{e}{16}X^2\tilde R +A'''\tfrac{i}{4}X^2 \bar
f^{\mu}\chi_{\mu},
\end{eqnarray}
where the curvature term\footnote{In the original version of this
paper the curvature term was induced from a shift in the spin
connection by $-\tfrac{1}{16}\ep_{\mu\alpha\beta}X^2$. This is
correct to order $(cov.der)^2$ but is in general not compatible with
BLG as the flat limit of the gauged theory.} is well-known to have
exactly the coefficient $-\tfrac{1}{16}$ in three dimensions so that
when added to the scalar kinetic term one obtains a locally scale
invariant expression.

Recalling the way the transformation rules for the gauge fields
$A_{\mu}$ and $B_{\mu}$ are obtained  we infer that both $\delta
A_{\mu}^{ab}$ and $\delta B_{\mu}^{ij}$ will pick up new terms in
the process of constructing the coupled theory. This is natural in
view of the fact that we work on-shell and that such terms are
expected to arise when auxiliary fields are eliminated. We start
from the following basic transformation rules without such terms
\begin{eqnarray}
\delta e_{\mu}{}^{\alpha}&=&i\bar\epsilon_g
\gamma^{\alpha}\chi_{\mu},\notag\\[1mm] \delta\chi_{\mu}&=&
\tilde D_{\mu}\epsilon_g, \notag\\[1mm]
 \delta B_{\mu}^{ij}&=&
-\tfrac{i}{2}\bar\epsilon_g\Gamma^{ij}\gamma_{\nu}\gamma_{\mu}f^{\nu}
, \notag\\[1mm] \delta X_{i}^a&=& i\epsilon_m \Gamma_i
\Psi^a,\notag\\[1mm] \delta\Psi_a&=&(\tilde D_{\mu}X^i_a -
i\hat A\bar\chi_{\mu} \Gamma^i \Psi_a) \gamma^{\mu} \Gamma^i
\epsilon_m+\tfrac{1}{6}X^i_b X^j_c X^k_d \Gamma^{ijk}\epsilon_m
f^{bcd}{}_a, \notag\\[1mm] \delta \tilde A_{\mu}{}^a{}_b&=&
i\bar\epsilon_m \gamma_{\mu} \Gamma^i X_c^i \Psi_d f^{cda}{}_b,
\end{eqnarray}
where $\epsilon_g$ and $\epsilon_m$ are the supersymmetry parameters
in the gravity and matter (BLG) sector, respectively. We are here
using different supersymmetry parameters in the two sectors since,
as we will see below, it will be necessary to rescale the
supersymmetry parameters relative each other for the Lagrangian to
be invariant.

As just mentioned, both $\delta\tilde A _{\mu}^{ab}$ and $\delta
B_{\mu}^{ij}$ will pick up a number of new terms as we proceed with
the calculation. By inspecting the possible terms we conclude
directly that these additional pieces will not contain any
derivatives $\tilde D_{\mu}$.  Some of these are (with
multiplicative constants and supersymmetry parameters to be
determined)
\begin{eqnarray}
\delta A _{\mu}^{ab}\vert_{new}=A_1\bar\chi_{\mu}\Gamma^{ij}\ep
X_a^iX_b^{j},
\end{eqnarray}
and
\begin{eqnarray}
\delta B_{\mu}^{ij}\vert_{new}=B_1\bar\Psi^a
\gamma_{\mu}\Gamma^{[i}\epsilon
X^{j]}_a+B_2\bar\chi_{\mu}\Gamma^{k[i}\ep
X_a^{j]}X_a^k+B_3\bar\Psi_a\Gamma^k\Gamma^{ij}\gamma_{\mu}\ep X_a^k.
\end{eqnarray}
We may find still others as we go through the proof of supersymmetry
at the $(\tilde D_{\mu})^2$ level. It is important to note in this
context that these new terms will not feed back into the proof of
supersymmetry at the order which we are working here, namely
$(\tilde D_{\mu})^2$. The proof that the lower order terms in $\de
L$ also cancel will, however, be affected.

The first step is to vary $L_{conf.sugra}+L_{BLG}^{cov}$ and keep
only the $(\tilde D_{\mu})^2$ terms that are not directly canceled,
along with all $(\tilde D_{\mu})^3$, in the pure supergravity case.
That is, we here use the fact that the supergravity sector is
invariant by itself as proved in the previous section. This means
that we can drop the torsion part which is not a derivative term.
When this is done it is possible to integrate by parts without
problems. We find
\begin{eqnarray}
&&\de L_{conf.sugra}\vert_{D^2}+\de L_{BLG}^{cov}\vert_{D^2}=
-\tfrac{1}{2}\delta(eg^{\mu\nu})D_{\mu}X_a^iD_{\nu}X_a^i-eD^{\mu}X_a^iD_{\mu}\delta
X_a^i\notag\\[1mm]
&&+ie\bar\Psi^a\gamma^{\mu}D_{\mu}\delta\Psi^a +e\delta
B_{\mu ij}\vert_{grav}(X_a^iD^{\mu}X_a^j)\notag\\[1mm] &&
+\tfrac{1}{2}\epsilon^{\mu\nu\rho}\delta
A_{\mu}^{ab}\vert_{BLG+new}\tilde
F_{\nu\rho}^{ab}+\epsilon^{\mu\nu\rho}\delta B_{\mu}^{ij}\vert_{new}
G_{\nu\rho}^{ij},
\end{eqnarray}
where the fourth term on the right hand side contributes to
$(D_{\mu})^2$ only if we insert the original supergravity variation
of $\delta B_{\mu}^{ij}$ as indicated. From now on we will not
include the last two terms proportional to the field strengths
$\tilde F_{\mu\nu}^{ab}$ and $G_{\mu\nu}^{ij}$ explicitly in our
expressions. When the variations of the potentials need to be
corrected we just have to recall their form from the above
expression.

Computing the above variation gives
\begin{eqnarray}
&&\de L_{conf.sugra}\vert_{D^2}+\de L_{BLG}^{cov}\vert_{D^2}=
-ieD_{\mu}X_a^iD_{\nu}X_a^i(\bar\chi^{\mu}\gamma^{\nu}\epsilon_g-
\tfrac{1}{2}g^{\mu\nu}\bar\chi_{\rho}\gamma^{\rho}\epsilon_g)\notag\\[1mm]
&&+\tfrac{i}{2}\epsilon^{\mu\nu\rho}
\bar\Psi^a\gamma_{\rho}\Gamma^i\epsilon_m G_{\mu\nu
ij}X_a^j+ie\bar\Psi^a\gamma^{\mu}\gamma^{\nu}\Gamma^iD_{\mu}\epsilon_m(D_{\nu}X_a^i)\notag\\[1mm]
&&+\tfrac{i}{2}\bar
f^{\nu}\gamma_{\mu}\gamma_{\nu}\Gamma^{ij}\epsilon_g(X_a^iD^{\mu}X_a^j),
\end{eqnarray}
where we find the first new contribution to the variation of $\delta
B_{\mu}^{ij}$. Choosing $B_1=-\tfrac{i}{2}$ and $\ep=\ep_m$ will
then remove the $G_{\mu\nu}$ term from this expression.

To eliminate some of the other terms we now add the first part  of
the supercurrent  $L_{supercurrent}\vert_{D^2}= Aie
\bar\chi_{\mu}\Gamma^i \gamma^{\nu}\gamma^{\mu}\Psi^a \tilde
 D_{\nu}X^{ia}$ (the other terms do not contribute to $(D_{\mu})^2$
 when varied under supersymmetry). $(D_{\mu})^2$ terms come from the
  variations $\de\chi_{\mu}$ and $\de\Psi_a$ which leave three terms
  (containing $\Gamma^{ij}$, $ \Gamma^{i}$ and no $\Gamma^{i}$'s)
 two of which cancel the first and third terms on the right hand
 side above provided $A\epsilon_g=\epsilon_m$ and
 $2A\epsilon_m=\epsilon_g$. Thus we conclude that
\begin{eqnarray}
\ep_m:=\ep,\,\,\,\ep_g=\pm \sqrt2\ep,\,\,\,A=\pm\tfrac{1}{\sqrt2},
\end{eqnarray}
where the sign of A will be chosen later.

The remaining terms are then
\begin{eqnarray}
\de
L\vert_{D^2}=-Ai\ep^{\mu\nu\rho}\bar\chi_{\mu}\Gamma^{ij}\ep_m(D_{\nu}X_a^iD_{\rho}X_a^j)
+\tfrac{i}{2}\bar
f^{\nu}\gamma_{\mu}\gamma_{\nu}\Gamma^{ij}\epsilon_g(X_a^iD^{\mu}X_a^j).
\end{eqnarray}
We now  add to $L$ the term
\begin{eqnarray}
A'i\epsilon^{\mu\nu\rho}\bar\chi_{\mu}\Gamma^{ij}\chi_{\nu}(X_a^i\tilde
D_{\rho}X_a^j),
\end{eqnarray}
since when $\chi_{\mu}$ is varied and the resulting expression
integrated by parts the term above proportional to $A$ is canceled
if we choose $A'=-\tfrac{1}{4}$. We also find new contributions to
the variations of $\delta\tilde A _{\mu}^{ab}$ and $\delta
B_{\mu}^{ij}$ corresponding to $A_1=2iA',\ep=\ep_g$ and
$B_2=iA',\ep=\ep_g$.

Due to a second cancelation, arising for $A'=-\tfrac{1}{4}$ (where from now
on we will choose the signs as $A=\tfrac{1}{\sqrt2}$,
$\ep_g=\sqrt2 \ep$), in the
previous step only one term remains at this stage, namely
\begin{eqnarray}
-\tfrac{i}{2}\bar
f^{\nu}\gamma_{\nu}\gamma^{\mu}\Gamma^{ij}\epsilon_m(X_a^iD_{\mu}X_a^j).
\end{eqnarray}
Thus also the term
\begin{eqnarray}
A''i\bar f^{\mu}\gamma_{\mu}\Gamma^i\Psi_aX_a^i
\end{eqnarray}
is needed, where the variation of both $\chi_{\mu}$ and $\Psi_a$
will produce $(D_{\mu})^2$ terms. All $\Gamma^{ij}$ terms are
eliminated by choosing $A''=\tfrac{1}{\sqrt2}$ and
$B_3=-A''\tfrac{i}{16}, \ep=\ep_g$. This leaves us with the
following variation (recalling that $R^{**}$ is a double density)
\begin{eqnarray}
\de
L\vert_{D^2}=-A''\tfrac{i}{4e}R^{**}\bar\ep_g\Gamma^i\Psi_aX_a^i+A''\tfrac{i}{2}(D_{\mu}X^2)(\bar
f^{\mu}\ep_m-\tfrac{1}{e}\ep^{\mu\nu\rho}\bar
f_{\nu}\gamma_{\rho}\ep_m).
\end{eqnarray}

Finally, we include the gravity term that is necessary to make the
scalar field kinetic term locally scale invariant (which fixes the
coefficient as given), that is,
\begin{eqnarray}
L_R=-\tfrac{e}{16}\tilde RX^2.
\end{eqnarray}
We will also need the associated  fermionic term
\begin{eqnarray}
L_{ferm}=A'''iX^2\bar f^{\mu}\chi_{\mu}.
\end{eqnarray}
The variation of the Ricci scalar term reads
\begin{eqnarray}
\de L_R\vert_{D^2}=\tfrac{i}{4e}R^{**}X_a^i\bar\ep_m\Gamma^i\Psi_a -
\tfrac{i}{8e}\bar\chi^{\mu}\gamma^{\nu}\ep_gR^{**}_{\mu\nu}X^2+
\tfrac{i}{4}\ep^{\mu\nu\rho}(D_{\mu}X^2)\bar f_{\nu}\gamma_{\rho}\ep_g.
\end{eqnarray}
Adding this to the last expression above for $\de L\vert_{D^2}$ we
see that the first terms in these two expressions cancel against
each other, as do the last terms, provided we use the fact that
$\ep_g=\sqrt2\ep_m$ as found above.

Thus, after including also the curvature scalar term we have
\begin{eqnarray}
\de L\vert_{D^2}=A''\tfrac{i}{2}(D_{\mu}X^2)\bar f^{\mu}\ep_m-
\tfrac{i}{8e}\bar\chi^{\mu}\gamma^{\nu}\ep_gR^{**}_{\mu\nu}X^2.
\end{eqnarray}
The final step is then to add the variation of the  fermionic term,
that is,
\begin{eqnarray}
\de L_{ferm}\vert_{D^2}=A'''iX^2\bar
f^{\mu}D_{\mu}\ep_g+A'''iX^2\bar\chi_{\mu}\de f^{\mu},
\end{eqnarray}
where the last term becomes, up to a $G_{\mu\nu}^{ij}$ field
strength term,
\begin{eqnarray}
A'''i\tfrac{1}{4e}\bar\chi^{\mu}\gamma^{\nu}\ep_gR^{**}_{\mu\nu}X^2.
\end{eqnarray}
(This can also be expressed in terms of the ordinary Ricci tensor as
\begin{eqnarray}
-A'''i\tfrac{e}{4}(\bar\ep_g\gamma_{\mu}\chi_{\nu}-
\tfrac{1}{2}g_{\mu\nu}\bar\ep_g\gamma^{\rho}\chi_{\rho})R^{\mu\nu}X^2,
\end{eqnarray}
where we used the relations
\begin{eqnarray}
R^{**}_{\mu\nu}=R_{\mu\nu}-\tfrac{1}{2}g_{\mu\nu}R,
\,\,\,\,R^{**}=-\tfrac{1}{2}R,
\end{eqnarray}
between the double dual $R^{**}_{\mu\nu}$ and the ordinary Ricci
tensor.)

The $G_{\mu\nu}^{ij}$ term mentioned in the previous paragraph
implies the following addition to $\delta B_{\mu}^{ij}$:
\begin{eqnarray}
\delta
B_{\mu}^{ij}\vert_{new:S}=\tfrac{i}{64}X^2\bar\ep_g\Gamma^{ij}\chi_{\mu},
\end{eqnarray}
which is just a special superconformal transformation with parameter
\begin{eqnarray}
\eta=\tfrac{1}{32}X^2\ep_g.
\end{eqnarray}
This indicates that also $\de\chi_{\mu}$ will pick up a special
superconformal piece:
\begin{eqnarray}
\de\chi_{\mu}\vert_S=\gamma_{\mu}\eta=\tfrac{1}{32}X^2\gamma_{\mu}\ep_g=-\tfrac{1}{16\sqrt2}X^2\gamma_{\mu}\ep.
\end{eqnarray}
As we will see below this will, in fact, not happen. However, another contribution to $\delta B_{\mu}^{ij}$ will
arise in the computation just below that will exactly double
the above special superconformal part of this transformation
rule\footnote{We thank Xiaoyong Chu for pointing out a sign
error in the first version of the paper.}.

Summing up the situation at this point, using what we know about the
various constants, we find that
\begin{eqnarray}
\de L\vert_{D^2}=\sqrt2 A'''i\bar f^{\mu}D_{\mu}\ep
X^2+\tfrac{i}{2\sqrt2}\bar f^{\mu}\ep D_{\mu}X^2- \tfrac{i}{4\sqrt2
e}(1-2A''')\bar\chi^{\mu}\gamma^{\nu}\ep R^{**}_{\mu\nu}X^2.
\end{eqnarray}

Thus if we choose $A'''=\tfrac{1}{4}$ the first two terms add and
the result can be integrated by parts to give
\begin{eqnarray}
 -\tfrac{i}{2\sqrt2}\bar \ep D_{\mu}f^{\mu} X^2= \tfrac{i}{8\sqrt2 e}\bar\chi^{\mu}\gamma^{\nu}\ep
R^{**}_{\mu\nu}X^2,
\end{eqnarray}
modulo another $G_{\mu\nu}^{ij}$ term, and hence we see that the
$(D_{\mu})^2$ terms vanish in the variation of the Lagrangian.

We now turn to the hatted coefficients in the Lagrangian given in
the beginning of this subsection. These can be determined as
follows. Consider first $\hat A$. This parameter is fixed by
demanding that the variation of $\Psi_a$ is
supercovariant\footnote{That is, if varied the right hand side of
$\delta \Psi_a$ must not give rise to derivatives on the
supersymmetry parameter.}, which gives $\hat A=\tfrac{1}{\sqrt2}$.

Turning to $\hat B$, we see that the variation of the dreibein in
the covariantized sixth order potential term is canceled by choosing
$\hat B=\tfrac{1}{6\sqrt2}$. The $\hat B$ term also gives rise to a
$X^6$ term containing
$\Gamma^{ijmn}(X^i_bX^j_cX^k_d)f^{abcd}(X^m_eX^n_fX^k_g)f^{aefg}$
which implies antisymmetry in $[bcef]$. However, this does not
immediately mean that the fundamental identity will set it to zero,
but by imposing $[abcf]$ on the fundamental identity (\ref{FI}) and
using its alternative  form given in (\ref{WFI}), that sets the left
hand side to zero, one finds that $f^{ab[cd}f^{ef]ag}=0$ which is
what we need.

The third, and last, parameter to be determined is $\hat C$. This we
do by relating the $\delta\chi_{\mu}$ variation of this term to the
two terms obtained by varying $\Psi_{a}\vert_{DX}$ in the $\hat B$
term and $\delta\Psi_a\vert_{X^3}$ in the supercurrent. We find
$\hat C=0$ due to a delicate cancelation.

We end this subsection by summarizing our results: Up to
terms\footnote{It might be that all terms of this kind are already
accounted for by the ones built into the covariant derivative in
which case the presented Lagrangian is the complete answer. Note
that terms like
$\bar\chi_{\mu}\Gamma^{ijkl}\chi^{\mu}\bar\Psi_{a}\Gamma^{ijkl}\Psi^{a}=0$
due to the chirality properties, and that
$\bar\chi_{\mu}\chi^{\mu}\bar\Psi_{a}\Psi^{a}$ is already present in
the supercurrent.} of order three or higher in $\chi_{\mu}$, the
Lagrangian reads
\begin{eqnarray}
L_{BLG}^{top}=&=&\frac{1}{2}\epsilon^{\mu\nu\rho}
Tr_{\alpha}(\tilde\omega_{\mu}\partial_{\nu}\tilde\omega_{\rho}+
\frac{2}{3}\tilde\omega_{\mu}\tilde\omega_{\nu}\tilde\omega_{\rho})
-\epsilon^{\mu\nu\rho}Tr_i
(B_{\mu}\partial_{\nu}B_{\rho}+\frac{2}{3}B_{\mu}B_{\nu}B_{\rho})
\nonumber
\end{eqnarray}
\begin{eqnarray}
&&-i e^{-1} \epsilon^{\alpha\mu\nu}\epsilon^{\beta\rho\sigma}(\tilde
D_{\mu}\bar{\chi}_{\nu}\gamma_{\beta}\gamma_{\alpha}\tilde
D_{\rho}\chi_{\sigma})\nonumber
\end{eqnarray}
\begin{eqnarray}
+e(-\tfrac{1}{2}g^{\mu\nu}\tilde D_{\mu}X^{ia} \tilde D_{\nu}X_{ia}
+\tfrac{i}{2}\bar\Psi^a \gamma^{\alpha}e_{\alpha}{}^{\mu}\tilde
D_{\mu}\Psi_a+L_{Yukawa}-V)+L_{CS(A)}\nonumber
\end{eqnarray}
\begin{eqnarray}
+ \tfrac{1}{\sqrt2}ie \bar\chi_{\mu}\Gamma^i
\gamma^{\nu}\gamma^{\mu}\Psi^a (\tilde
 D_{\nu}X^{ia}-\tfrac{i}{2\sqrt2}\bar\chi_{\nu}\Gamma^i \Psi^a)\nonumber
 \end{eqnarray}
\begin{eqnarray}
 -\tfrac{1}{6\sqrt2}ie\bar\chi_{\mu}\gamma^{\mu}\Gamma^{ijk}\Psi_a(X^i_bX^j_cX^k_d)f^{abcd}\nonumber
\end{eqnarray}
\begin{eqnarray}
-\tfrac{i}{4}\epsilon^{\mu\nu\rho}\bar\chi_{\mu}\Gamma^{ij}\chi_{\nu}(X_a^i\tilde
D_{\rho}X_a^j)+ \tfrac{i}{\sqrt2}\bar
f^{\mu}\Gamma^i\gamma_{\mu}\Psi_a X_a^i\nonumber
\end{eqnarray}
\begin{eqnarray}
-\tfrac{e}{16}X^2\tilde R +\tfrac{i}{4}X^2 \bar
f^{\mu}\chi_{\mu}\,,
\end{eqnarray}
and the transformation rules are
\begin{eqnarray}
\delta e_{\mu}{}^{\alpha}&=&i\sqrt2\bar\epsilon
\gamma^{\alpha}\chi_{\mu}\,,\notag\\[1mm]
 \delta\chi_{\mu}&=& \sqrt2\tilde
D_{\mu}\epsilon, \notag\\[1mm]
 \delta B_{\mu}^{ij}&=&
-\tfrac{i}{\sqrt2
e}\bar\epsilon\Gamma^{ij}\gamma_{\nu}\gamma_{\mu}f^{\nu}
-\tfrac{i}{2}\bar\Psi_a \gamma_{\mu}\Gamma^{[i}\epsilon
X^{j]}_a-\tfrac{i}{2\sqrt2}\bar\chi_{\mu}\Gamma^{k[i}\ep
X_a^{j]}X_a^k-\tfrac{i}{16}\bar\Psi_a\Gamma^k\Gamma^{ij}\gamma_{\mu}\ep
X_a^k\notag\\[1mm]
&&+\tfrac{i}{16\sqrt2} \bar\epsilon \Gamma^{ij}\chi_\mu X^2
 , \notag\\[1mm]
 \delta X_{i}^a&=& i\bar\epsilon \Gamma_i \Psi^a,\notag\\[1mm]
\delta\Psi_a&=&(\tilde D_{\mu}X^i_a -
\tfrac{i}{\sqrt2}\bar\chi_{\mu} \Gamma^i \Psi_a) \gamma^{\mu}
\Gamma^i \epsilon+\tfrac{1}{6}X^i_b X^j_c X^k_d
\Gamma^{ijk}\epsilon f^{bcd}{}_a, \notag\\[1mm] \delta \tilde
A_{\mu}{}^a{}_b&=& i\bar\epsilon \gamma_{\mu} \Gamma^i X_c^i
\Psi_d
f^{cda}{}_b-\tfrac{i}{\sqrt2}\bar\chi_{\mu}\Gamma^{ij}\ep
X_c^iX_d^{j}f^{cda}{}_b.
\end{eqnarray}
What remains to be checked are the terms in $\de L$ that are linear
in the covariant derivative or independent of them. We hope to
present this final step of the proof elsewhere.

\section{Conclusions and comments}\label{concl}

In this paper we have constructed the $\mathcal N=8$ conformal
supergravity theory in three dimensions that seems to be the proper
theory to couple to the $\mathcal N=8$ BLG theory believed to
describe two M2 branes at the IR conformal fix-point. The $\mathcal
N=8$ conformal supergravity theory consists on-shell of just three
Chern-Simons type terms one for each of the gauge fields, the spin
connection (in second order form), the Rarita-Schwinger and $SO(8)$
R-symmetry gauge fields. This theory should be possible to couple to
matter in the form of the BLG theory which is a rather lengthy
operation to do in full detail. The construction carried out in this
paper, relying on the cancelation in $\delta L$ of terms containing
two or three covariant derivatives, generates the complete
Lagrangian apart from some fermionic interaction terms.

There are several aspects of the gauged BLG theory that might be of
interest. Free Chern-Simons gauge theories are really topological
theories whose symmetries become reduced to superconformal ones when
coupled to each other (as in the supergravity sector) or to
conformal matter (as in the BLG sector) although the gravity sector
is probably somewhat more intricate. In any case, what seems to be a
general feature is that the various curvatures are heavily
restricted, or even determined, by the field equations. For pure
Chern-Simons gravity this is discussed for instance in
\cite{Horne:1988jf}. A Lagrangian based on a second order spin
connection leads to the equation of motion
\begin{equation}
D_{[\mu}W_{\nu]\rho}=0, \,\,\,
W_{\mu\nu}=R_{\mu\nu}-\tfrac{1}{4}g_{\mu\nu}R,
\end{equation}
which is known to be the condition for conformal flatness in three
dimensions. This equation will be modified by source terms
constructed from the other fields appearing in the theory.

Another well-known property of the BLG theory is that it allows for
the introduction of a level $k$
\cite{VanRaamsdonk:2008ft,Lambert:2008et,Distler:2008mk} which can
be seen by using structure constants of the form
\begin{equation}
f^{abcd}=\frac{2\pi}{k}\epsilon^{abcd}.
\end{equation}
Then reabsorbing one such factor $\frac{2\pi}{k}$ into the gauge
field in the BLG theory produces the level $k$ theory where $k$ is
an integer for topological reasons. If this is done in the Van
Raamsdonk version \cite{VanRaamsdonk:2008ft} one finds the standard
level $(k,-k)$ theory discussed more generally in
\cite{Aharony:2008ug}. Interestingly enough the coupling to
superconformal gravity discussed in this paper introduces yet
another level parameter which is also quantized as explained in
\cite{Horne:1988jf}. There seems, however, to be room for only one
new such parameter in the $\mathcal N=8$ superconformal case since
the extra Chern-Simons terms are connected by the various local
symmetries. It is perhaps interesting to note in this context that
the Chern-Simons term for the R-symmetry field $B_{\mu}^{ij}$ gets
here an unconventional normalization (being twice the standard one).

This last issue relates also to the question of invariance under
parity for the gauged BLG theory. The pure BLG theory is saved by
the fact that in the $SU(2)\times SU(2)$ formulation in
\cite{VanRaamsdonk:2008ft,Lambert:2008et,Distler:2008mk} the two
gauge groups are interchanged by a parity transformation. This
option seems not to be available in the superconformal gravity
sector as formulated here.

Apart from the original $\mathcal N=8$ BLG theory there are a number
of other versions of superconformal M2 brane theories with less
supersymmetry but able to describe more general stacks of branes.
Following \cite{VanRaamsdonk:2008ft}, the authors of
\cite{Aharony:2008ug} (see also \cite{Benna:2008zy,Bandres:2008ry})
used a construction with fields in the bi-fundamental representation
of $U(N) \times U(N)$ relevant for stacks with $N$ branes. So far,
however, this ABJM theory exists only with 6 supersymmetries which,
however, may get enhanced to 8 for level $k=1,2$ if monopole
operators are introduced \cite{Aharony:2008ug,Klebanov:2008vq}. In
such a context infinite dimensional algebraic structures will
probably play a role. An example of such a structure, related to
generalized Jordan triple products, has recently been suggested to
arise in BLG/ABJM theories \cite{Nilsson:2008kq}. Here we have not
made an attempt to couple the ABJM theory to $\mathcal N=6$
superconformal gravity but it should be possible and follow the same
lines as those used in this paper. Another method that might be
useful in this context is the embedding tensor technique already
applied to similar problems, for instance, in
\cite{Bergshoeff:2008bh}.

\acknowledgments

We would like to thank Martin Cederwall, Xiaoyong Chu, Gabriele Ferretti, Andreas
Gustavsson, George Papadopoulos, Christoffer Petersson and Arkady
Tseytlin for discussions. The work is partly funded by the Swedish
Research Council.

\appendix

\section{Fierz bases}

 The Fierz basis used in the proof of supersymmetry
in the main text is based on expressions of the form
$(\bar\epsilon...\chi_{\mu})(\bar f_{\nu}...f_{\rho})$ where the
dots refers to either an antisymmetric three-dimensional charge
conjugation matrix or to a product of it with a three-dimensional
gamma which is symmetric. Thus these expressions have three, four or
five free indices that need to be contracted by deltas or
Levi-Civita symbols. There are thus twelve index structures :
\begin{eqnarray}
&(-)&(\bar\epsilon\chi_{\mu})(\bar
f_{\nu}f_{\rho})\epsilon^{\mu\nu\rho}=0,\cr
&(-)&(\bar\epsilon\chi_{\alpha})(\bar
f^{\beta}\gamma^{\alpha}f_{\beta})=0,\cr
&(1)&(\bar\epsilon\chi_{\alpha})(\bar
f^{\alpha}\gamma^{\beta}f_{\beta}),\cr
&(2)&(\bar\epsilon\gamma^{\alpha}\chi_{\alpha})(\bar
f^{\beta}f_{\beta}),\cr
&(3)&(\bar\epsilon\gamma_{\alpha}\chi_{\beta})(\bar
f^{\alpha}f^{\beta}),\cr
&(4)&(\bar\epsilon\gamma^{\alpha}\chi_{\alpha})(\bar
f_{\mu}\gamma_{\nu}f_{\rho})\epsilon^{\mu\nu\rho},\cr
&(5)&(\bar\epsilon\gamma_{\mu}\chi_{\nu})(\bar
f_{\rho}\gamma^{\alpha}f_{\alpha})\epsilon^{\mu\nu\rho},\cr
&(6)&(\bar\epsilon\gamma^{\alpha}\chi_{\mu})(\bar
f_{\alpha}\gamma_{\nu}f_{\rho})\epsilon^{\mu\nu\rho},\cr
&(7)&(\bar\epsilon\gamma^{\alpha}\chi_{\mu})(\bar
f_{\nu}\gamma_{\alpha}f_{\rho})\epsilon^{\mu\nu\rho},\cr
&(8)&(\bar\epsilon\gamma^{\mu}\chi_{\alpha})(\bar
f^{\alpha}\gamma_{\nu}f_{\rho})\epsilon^{\mu\nu\rho},\cr
&(9)&(\bar\epsilon\gamma_{\mu}\chi_{\alpha})(\bar
f_{\nu}\gamma^{\alpha}f_{\rho})\epsilon^{\mu\nu\rho},\cr
&(-)&(\bar\epsilon\gamma_{\mu}\chi_{\nu})(\bar
f_{\alpha}\gamma^{\rho}f_{\alpha})\epsilon^{\mu\nu\rho}=0.
\end{eqnarray}

Of the ten non-zero ones we can easily (by cycling the three indices
on the epsilon tensor together with on of the contracted indices)
find three relations involving the expressions (4) to (9):
\begin{eqnarray}
2 \cdot(6)=(4)+(9),\,\,\, 2\cdot(5)=(7)-(9),\,\,\, (4)=2
\cdot(6)-(7).
\end{eqnarray}
We will choose as an independent set of expressions (1),(2),(3),
(4), (6), and (8), which means that $(9)=2\cdot(8)-(4)$,
$(7)=2\cdot(6)-(4)$, and $(5)=(6)-(8)$.

We may also relate this basis to expressions that appear frequently
in the Lagrangian:
\begin{eqnarray}
(\hat4)&:=&(\bar\epsilon\gamma^{\alpha}\chi_{\alpha})(\bar
f_{\beta}\gamma^{\gamma}\gamma^{\beta}f_{\gamma})\cr (\hat
6)&:=&(\bar\epsilon\gamma^{\alpha}\chi_{\beta})(\bar
f_{\alpha}\gamma^{\gamma}\gamma^{\beta}f_{\gamma}),\cr
(\hat8)&:=&(\bar\epsilon\gamma^{\alpha}\chi_{\beta})(\bar
f_{\beta}\gamma^{\gamma}\gamma^{\alpha}f_{\gamma}).
\end{eqnarray}
Expressing these in the basis specified above gives
\begin{eqnarray}
(\hat4)=(4)+(2), \,\,(\hat6)=(6)+(3),\,\,(\hat8)=(8)+(3).
\end{eqnarray}

When the $SO(8)$ $\Gamma$-matrices are introduced into the Fierz
identity the same basis can be used by inserting $\Gamma$'s into
both factors. For $\Gamma^{ijkl}$ the basis is exactly the same as
the one above while for $\Gamma^{ij}$ some other elements are set to
zero by symmetry
\begin{eqnarray}
&(1')&(\bar\epsilon\Gamma^{ij}\chi_{\mu})(\bar
f_{\nu}\Gamma^{ij}f_{\rho})\epsilon^{\mu\nu\rho},\cr
&(2')&(\bar\epsilon\Gamma^{ij}\chi_{\alpha})(\bar
f^{\beta}\gamma^{\alpha}\Gamma^{ij}f_{\beta}),\cr
&(3')&(\bar\epsilon\Gamma^{ij}\chi_{\alpha})(\bar
f^{\alpha}\gamma^{\beta}\Gamma^{ij}f_{\beta}),\cr
&(-)&(\bar\epsilon\Gamma^{ij}\gamma^{\alpha}\chi_{\alpha})(\bar
f^{\beta}f_{\beta})=0,\cr
&(4')&(\bar\epsilon\gamma_{\alpha}\Gamma^{ij}\chi_{\beta})(\bar
f^{\alpha}\Gamma^{ij}f^{\beta}),\cr
&(-)&(\bar\epsilon\gamma^{\alpha}\Gamma^{ij}\chi_{\alpha})(\bar
f_{\mu}\gamma_{\nu}\Gamma^{ij}f_{\rho})\epsilon^{\mu\nu\rho}=0,\cr
&(5')&(\bar\epsilon\gamma_{\mu}\Gamma^{ij}\chi_{\nu})(\bar
f_{\rho}\gamma^{\alpha}\Gamma^{ij}f_{\alpha})\epsilon^{\mu\nu\rho},\cr
&(6')&(\bar\epsilon\gamma^{\alpha}\Gamma^{ij}\chi_{\mu})(\bar
f_{\alpha}\gamma_{\nu}\Gamma^{ij}f_{\rho})\epsilon^{\mu\nu\rho},\cr
&(-)&(\bar\epsilon\gamma^{\alpha}\Gamma^{ij}\chi_{\mu})(\bar
f_{\nu}\gamma_{\alpha}\Gamma^{ij}f_{\rho})\epsilon^{\mu\nu\rho}=0,\cr
&(7')&(\bar\epsilon\gamma^{\mu}\Gamma^{ij}\chi_{\alpha})(\bar
f^{\alpha}\gamma_{\nu}\Gamma^{ij}f_{\rho})\epsilon^{\mu\nu\rho},\cr
&(-)&(\bar\epsilon\gamma_{\mu}\Gamma^{ij}\chi_{\alpha})(\bar
f_{\nu}\gamma^{\alpha}\Gamma^{ij}f_{\rho})\epsilon^{\mu\nu\rho}=0,\cr
&(8')&(\bar\epsilon\gamma_{\mu}\Gamma^{ij}\chi_{\nu})(\bar
f_{\alpha}\gamma^{\rho}\Gamma^{ij}f_{\alpha})\epsilon^{\mu\nu\rho},
\end{eqnarray}
and the set of independent basis elements can be chosen as
$(1'),(2'),(3'),(4'), (5')$ and $(7')$.

\end{document}